\documentclass[12pt]{article}

\usepackage{scicite}

\usepackage{times}

\usepackage{psfig,epsfig}

\newenvironment{sciabstract}{%
\begin{quote} \bf}
{\end{quote}}

\newcounter{lastnote}
\newenvironment{scilastnote}{%
\setcounter{lastnote}{\value{enumiv}}%
\addtocounter{lastnote}{+1}%
\begin{list}%
{\arabic{lastnote}.}
{\setlength{\leftmargin}{.22in}}
{\setlength{\labelsep}{.5em}}}
{\end{list}}

\title{Producing ultra-strong magnetic fields in neutron star mergers}

\author{D.J. Price$^{1}$ and S. Rosswog$^{2\ast}$ \\
\normalsize{$^{1}$School of Physics, University of Exeter, }\\
\normalsize{Stocker Road, Exeter EX4 4QL, United Kingdom}\\
\normalsize{$^{2}$School of Engineering and Science, International University Bremen,}\\
\normalsize{Campus Ring 1, 28759 Bremen, Germany }\\
\normalsize{$^\ast$To whom correspondence should be addressed; E-mail: s.rosswog@iu-bremen.de}
}

\date{}

\begin{document} 


\baselineskip24pt


\maketitle 


\begin{sciabstract}
%
%
We report an extremely rapid mechanism
for magnetic field amplification during the merger of a binary neutron
star system. This has implications for the production of the short
class of Gamma-Ray Bursts, which recent
observations suggest may originate in
such mergers. In detailed
magnetohydrodynamic simulations of the merger process, the fields are
amplified via Kelvin-Helmholtz instabilities beyond magnetar
field strength and may therefore represent the strongest magnetic fields
in the Universe. The amplification occurs in the shear layer which forms
between
the neutron stars and on a time scale of only 1 millisecond, i.e. long before
the remnant can collapse into a black hole.
\end{sciabstract}




The orbital decay of a neutron star binary system due to the emission of
gravitational waves is one of the prime targets of gravitational
wave detectors such as LIGO \cite{ligo92} or GEO600 \cite{geo600}. Moreover, the long-suspected connection of neutron star binaries to Gamma-Ray Bursts (GRBs), the
most luminous explosions in the universe, has received solid support from the first detections of afterglows from the short class of GRBs\cite{bloom06,berger05,barthelmy05}: unlike their long-duration cousins (which are associated with the deaths of massive stars) short GRBs occur systematically at lower redshifts, both in galaxies with and without star formation and are not accompanied
by a supernova explosion. The millisecond variability observed in the lightcurves of short GRBs suggests that a compact object, either a neutron star or a stellar mass black hole acts as the central engine.

 The observed cosmological distances imply that large energies are involved and therefore, to avoid the so-called ``compactness problem'' \cite{piran05}, relativistic outflows with Lorentz-factors of several hundreds are required. To reach such extreme velocities a large amount of energy has to be deposited per rest mass, for example by the annihilation of neutrino-antineutrino pairs, $\bar{\nu}_i+\nu_i \rightarrow
e+e^+$, or via magnetic mechanisms \cite{eichler89,narayan92}. Therefore, strong magnetic fields have been suggested as being important in producing GRBs \cite{usov92,duncan92,meszaros97,kluz98,lyutikov03}, but the question of what field strengths can actually be reached in a merger remnant before it collapses to a black hole, has so far remained unanswered. Recently, a very energetic giant flare from the magnetar SGR 1806-20 has been
observed\cite{hurley05,palmer05}. Would it have been further away, but within
40 Mpc, its initial spike would -both on grounds of duration and spectrum-
have been interpreted as a short gamma-ray burst. The lack of excess events
from the direction of the Virgo cluster, however, suggests that only a small
portion of previously observed short bursts could have been giant magnetar
flares. Nevertheless, the similarity in physical properties may point to a
common or similar mechanism behind both phenomena.

 Although computer simulations of binary neutron star mergers have reached a good degree of 
realism\cite{ruffert01,rosswog02,rosswog03a,rosswog03b}, none has so far been able to take magnetic fields into account, primarily due to the numerical challenge posed by simulating even the hydrodynamics of the merger process. Here we present global neutron star merger simulations that follow the evolution of the magnetic field. Our main result is that the existing neutron star magnetic fields ($10^{12}$G) become
amplified by several orders of magnitude within the first millisecond after the merger, which is long before the collapse to a black hole can proceed. Our robust lower limit on the field that can be reached is
$2\times10^{15}$G, but is it highly probable that much stronger fields are realized in
nature.

%
%

 Our simulations are three-dimensional computer simulations of 
two neutron stars that coalesce due to the emission of gravitational waves. The
equations of hydrodynamics are solved with a Lagrangian 
particle scheme (Smoothed Particle Hydrodynamics, SPH) [for a review see \cite{monaghan05}] that is coupled to a  
temperature- and composition-dependent nuclear equation of
state\cite{shen98,rosswog02}. We include the effects of cooling and the change in matter composition due to neutrino-producing weak interactions. As the debris material covers the full range from completely opaque to completely transparent to neutrinos, we have to incorporate opacity effects. Thus on an additional grid, we calculate for each fluid parcel the opacities for each neutrino species and take them into account in the emission process \cite{rosswog03a}. The Newtonian self-gravity of the neutron star fluid is evaluated efficiently using a binary tree algorithm. In addition we apply forces that emerge due to the emission of gravitational waves \cite{rosswog02} - these are the forces that drive the binary towards coalescence. The new physics employed in these simulations is the inclusion of magnetic fields. To ensure the robustness of our results we apply two different methods: one using a recently developed algorithm
for ``Smoothed Particle Magnetohydrodynamics'' \cite{price05} and one using a method where the magnetic field is calculated from the so-called ``Euler potentials'', $\alpha$ and $\beta$ that are advected with each fluid particle\cite{stern70}. The magnetic field is calculated from these potentials according to
\begin{equation}
{\bf B} = \nabla\alpha \times \nabla\beta.
\label{eq:euler}
\end{equation}
This prescription has the advantage that the divergence constraint (``no monopoles condition'') on the magnetic field is satisfied by construction. Apart from this difference, both methods yield similar results. The computational costs are dominated by the calculation of self-gravity -- the costs for the magnetic fields, the equation of state and the neutrino physics are negligible by comparison.


Our initial neutron stars are cold and have masses of
1.4 solar masses (M$_{\odot}$) each. The two stars are placed at an initial separation of 48 km with velocities corresponding to a circular orbit around their common centre of mass. As the inspiral dynamics only allows for a short time of tidal interaction, the neutron stars cannot be spun
up substantially\cite{bildsten92}, therefore we start our
calculations with non-spinning neutron stars. We choose uniform magnetic
fields of the typical strength of neutron stars ($10^{12}$ Gauss) as initial
conditions (note that we are modelling the interior fields, not the exterior dipole-like fields). The field orientation is parallel to the orbital angular momentum
for one star and opposite for the other. Test calculations with different
orientations yield maximum field strengths that are very similar.

The global dynamical evolution is shown (Fig. 1 and as an animation in the supporting online material): the two stars merge into
a single object within about one orbital period ($\sim 2$ms). Subsequently, excess angular
momentum is transported outward in spiral arms which quickly spread into a
thick accretion disk around the central object. When the
stars come into contact a shear interface forms, across which the tangential
velocity exhibits a jump (Fig. 2). In such a shear layer even infinitesimally small perturbations are unstable to the Kelvin-Helmholtz (KH) instability and will grow, in this case causing the interface to curl up into vortex rolls. An analogous effect occurs when wind blows across the surface of a lake, curling up the surface into waves. The initial growth rate, $\sigma_{\lambda}$, of the KH instability may be calculated analytically for the inviscid, incompressible case (both properties are good approximations for neutron stars). In the linear regime the growth rate is $\sigma_{\lambda}= \pi v/\lambda$ \cite{landaulifshitzfm}, where $\lambda$ is the wavelength of the growing mode and $v$ the velocity across the jump. As the
shortest modes grow fastest, the
numerical results can only be lower limits on the growth realized in
nature. Inserting the smallest length that we can numerically resolve for $\lambda$ and the simulation value for $v$ into the above
equation yields growth rates consistent with the simulations.

In all cases we find that the field in the vortex rolls is amplified
within $\sim$ 1 millisecond by orders of magnitude (Figs. 3 and 4). This time
scale should be 
compared to the 50-100 milliseconds which are estimated\cite{shibata05} for
the newly formed, differentially rotating central object to collapse into a
black hole. The high field strength material produced in the shear instability between the stars is subsequently advected with the matter to cover the surface of the central merger remnant (Fig. 1). 

 As the length scales we can resolve numerically are larger than the physical lengths that will trigger the KH instability in nature, our numerical results represent robust lower limits on the true magnetic field strengths. This is demonstrated by our numerical resolution study (Fig. 4). Each
time we double the numerical resolution (increase the particle
number by a factor of 10), the peak field strength increases by a factor of a
few. The highest numerical resolution that we can currently afford ($2 \times
10^6$ fluid particles), yields a field strength beyond $2 \times 10^{15}$
Gauss, i.e. larger than the largest magnetic fields that have been observed in
magnetars. The growth is likely to saturate when the magnetic field becomes strong enough to feed back on the fluid or alternatively when the field becomes buoyant. In either case the field strength reached will be comparable to the equipartition field strength (here $10^{16}$-$10^{18}$ G), where the magnetic pressure becomes comparable to the gas pressure.

 While we can only speculate about the field
strengths that will be actually produced in nature, it is clear that the strong magnetic
fields that have been conjectured in earlier work, occur naturally in the initial
shear phase even before possible dynamo mechanisms could have set in. If a fraction $\epsilon$ of the rotational energy of the central object of the remnant, $8\times 10^{52}$ erg, is channelled into the magnetic field, the field strength averaged over the central object will be $B = 1.2\times 10^{17} {\rm G} (\epsilon/0.1)^{1/2}(E_{kin}/8\times 10^{52} {\rm erg})^{1/2} (15 {\rm km}/R_{co})^{3/2}$, where $R_{co}$ is the radius of the central object. Locally, the field strength could be even higher. Near equipartition matter blobs in high field pockets (such as the vortices seen in Figure 3) will become buoyant, float up and and produce a relativistic blast as they break through the surface of the central object \cite{kluz98}. This could be the variable, relativistic outflow that is required to produce a GRB far from the central engine. In this case the millisecond substructures would be determined by the fluid instabilities in
the central object, but the overall duration would be set by the time it takes
the central object to collapse or to consume its rotational energy. If we use the magnetic dipole spin down time as an order of magnitude estimate and insert typical numbers from the simulation, we find
\begin{equation}
\tau = 0.14 {\rm s} \left(\frac{B}{10^{17}G}\right)^{-2} \left(\frac{P}{1.5 {\rm ms}}\right)^{2} \left(\frac{15 {\rm km}}{R_{co}}\right)^{4} \left(\frac{M_{co}}{2.5 M_{\odot}}\right),
\end{equation}
where $P$ is the rotational period and $M_{co}$ is the mass of the central object. This timescale is close to the typical duration of a short GRB. Note also that the high-field strength matter is transported by the fluid motion to the remnant surface. The sudden
appearance of $10^{17}$-G-material at the surface of the neutron-star-like
central object will very plausibly launch magnetized blasts similar to those described in \cite{lyutikov03b}. Similar processes involving buoyant magnetic fields, although at lower field strengths ($\sim 10^{14}-10^{16}$G), may also be at work in the accretion torus. It is also worth pointing out the somewhat speculative possibility that such a merger could produce magnetars.

Recent calculations\cite{rosswog03c,aloy05} in the GRB context have shown that the deposition of
thermal energy above accretion disks, for example from neutrino annihilation, can drive
relativistic outflows. Such outflows can be narrowly collimated, but in general a large spread in energies and opening angles depending on the specifics of the merging system is expected.
In the light of the above presented results, it is hard to see how a signal
from the strong magnetic fields -on top of 
the neutrino-annihilation driven outflows- can be avoided. 

 At the neutrino luminosities produced in the merger ($> 10^{52}$ erg/s), neutrinos will, as in the case of newborn neutron stars, drive a strong baryonic wind \cite{dsw86}. This material poses
a potential threat to the emergence of the required ultra-relativistic
outflow. The central object is rather hot ($20-25$ MeV, where $1$ Mev $ = 1.16\times 10^{11}K$), but very opaque to
      neutrinos. It therefore only contributes moderately to the total
      neutrino luminosity, which is dominated by the inner shock-heated torus
      regions, where we expect the most of the wind material to come from. Directly after
      the merger the environment is of very low density and rising
      magnetic bubbles will, via magnetic pressure, help to keep the region
      above the central object relatively clean of baryons. But as the neutrino
      luminosity rises and the continuously braked central object takes longer
      and longer to reach buoyancy field strength, it will become
      increasingly difficult to launch relativistic outflows. The interaction
      between such magnetic bubbles and a baryonic wind will be very
      complicated and whether relativistic outflow develops or not may depend
      on the details of the merging system. The estimated double neutron star merger rate ranges from about $4$ to $220\times 10^{-6}$ per year and galaxy \cite{kalogera04b}, and is thus comfortably two orders of magnitude larger than the rate required to explain short GRBs. Thus, even allowing for beaming and for a fraction of systems that could possibly fail to provide the right conditions (instead producing baryon-loaded X-ray or UV-flashes), the merger rate is still large enough to explain the observed short GRBs.
      
The two mechanisms -neutrino annihilation and magnetic processes- will show
a different temporal evolution and they
will also differ in the energies they can provide the burst with. The torus that dominates the neutrino emission takes a few dynamical time
scales to form (in our simulations the neutrino luminosity peaks
about 30 milliseconds after the stars have come in contact). The initial
amplification of the magnetic field occurs on a much shorter time
scale. Therefore, we expect the very early prompt emission to come from the
magnetic field alone. The outflows driven by neutrino-annihilation contain typically
$10^{48}$ ergs \cite{rosswog03b}, magnetic mechanisms could
easily provide $10^{51}$ ergs or more\cite{narayan92,rosswog03b}. Very energetic short bursts would
therefore have to be attributed to magnetic mechanisms. In any case, short GRBs that arise from the merger of magnetized neutron stars will exhibit a large intrinsic diversity, a
very complex temporal behavior and their observed properties will drastically
depend on their orientation relative to the line of sight.



\bibliographystyle{Science}
\bibliography{nsns}
\begin{scilastnote}
\item It is a pleasure to thank Marcus Br\"uggen, Matthias H\"oft, Joachim Vogt, Richard West and Matthew Bate for helpful discussions and comments. DJP is supported by a PPARC postdoctoral research fellowship. We thank the referees for their insightful comments.
\end{scilastnote}

\newpage
\begin{figure}
 \begin{center}
 \psfig{file=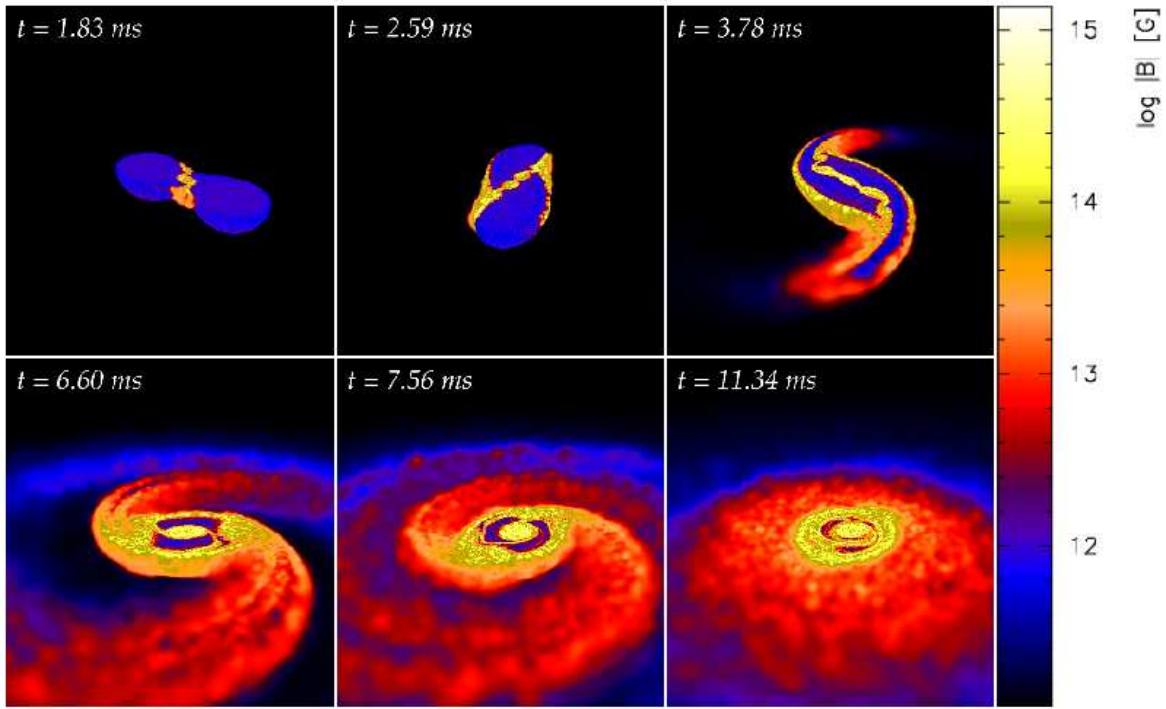,height=\textwidth,angle=-90}

\caption{Snapshots (left to right, top to bottom) of the coalescence of
      two magnetized neutron stars, showing magnetic field strengths in the material at and below the orbital plane. Dimensions in each panel are $\sim 140$ km from left to right. The stars move gradually towards each other and then merge in a ``plunging phase'' within about one
      orbital period ($\sim 2$ ms; first two snapshots). This object sheds mass into spiral
      arms that are subsequently wrapped around the central object (snapshots three to five) to form a hot torus (last snapshot).  The magnetic field is amplified in the shear instability between the stars and subsequently advected with the matter to cover the surface of the central merger remnant. }
    \label{fig:price_rosswog:1}
  \end{center}
\end{figure}

\newpage

\begin{figure}
 \begin{center}
\psfig{file=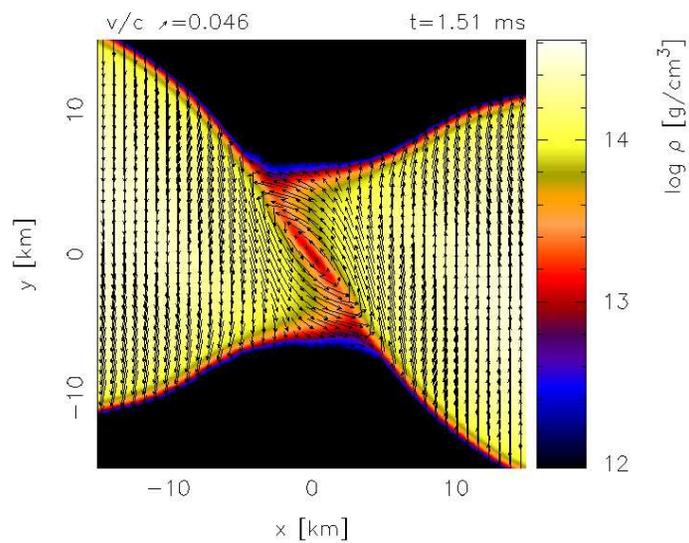,width=8cm,angle=-90}
    \caption{Density and velocity field in the orbital plane at
the moment when the stars come into contact (t=1.51 ms). A shear interface
forms between the stars, across which the tangential velocity exhibits a large
  jump. This interface is unstable to the Kelvin-Helmholtz instability and
  will curl up into vortex rolls (see next Figure).}
    \label{fig:price_rosswog:2}
  \end{center}
\end{figure}

\begin{figure}
 \begin{center}
\psfig{file=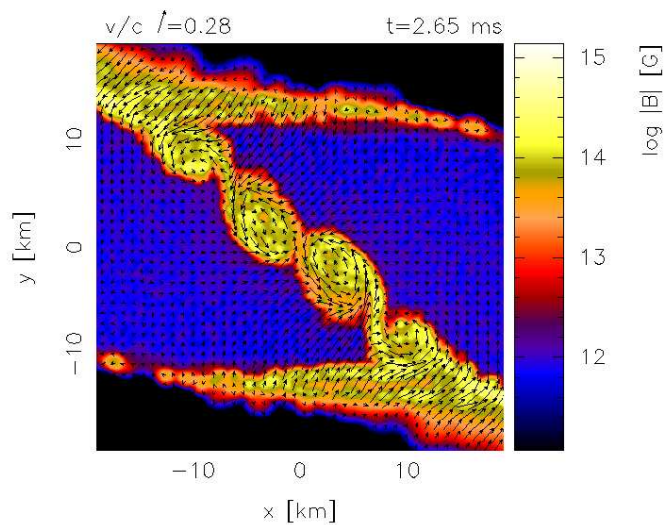,width=8cm,angle=-90}
    \caption{Close-up of the central regions at t=2.65 milliseconds. The colour coding shows strength of the magnetic field whilst the arrows show the fluid velocity in the corotating frame (that is, with the dominant orbital velocity component removed). 
    The shear interface shown in Figure 2 can be seen to have curled up into vortex rolls. In these
      vortices the field is strongly amplified to strengths exceeding $10^{15}$G. High-field material that has
      passed through these vortex rolls is subsequently spread across the
      surface of the central merger remnant (see first three panels of Figure 1).}
    \label{fig:price_rosswog:3}
 \end{center}
\end{figure}

\begin{figure}
 \begin{center}
\psfig{file=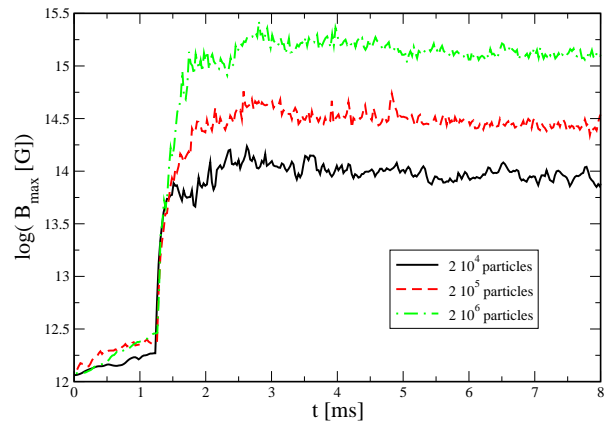,width=7cm,angle=-90}
    \caption{Maximum magnetic field strength as a function of the number
      of fluid particles. All three runs are identical apart from the numerical resolution. The maximum field strength of the best resolved run,
      $2\times 10^{15}$ G, is a strict lower limit on the magnetic field that
      can be reached in a neutron star merger.}
    \label{fig:price_rosswog:4}
  \end{center}
\end{figure}

\end{document}